\documentclass[aps,12pt,nofootinbib,superscriptaddress,preprint]{revtex4}
\pdfoutput=1
\usepackage{textcomp}
\usepackage{amssymb}
\usepackage{anysize}
\usepackage{bold-extra}
\usepackage{enumerate}
\usepackage{graphicx}
\usepackage{lscape}
\usepackage{mathrsfs}
\usepackage{multirow}
\usepackage[final]{pdfpages}
\usepackage{rotating}
\usepackage{subfig}
\usepackage{url}
\usepackage{wrapfig}
\usepackage{amsmath}
\usepackage{color}
\usepackage{fancyhdr}
\def\bea{\begin{eqnarray}}
\def\eea{\end{eqnarray}}
\def\ba{\begin{eqnarray}}
\def\ea{\end{eqnarray}}
\def\be{\begin{equation}}
\def\ee{\end{equation}}
\def\beq{\begin{equation}}
\def\eeq{\end{equation}}

\def\gsim{\mbox{\raisebox{-.6ex}{~$\stackrel{>}{\sim}$~}}}

\begin{document}
%\rhead{CERN-PH-TH/2011-261 \\ UCSD/PTH 11-18}
\preprint{CERN-PH-TH/2011-261}
\preprint{UCSD/PTH 11-18}
\vspace{20pt}
\title{EWPD Constraints on Flavor Symmetric Vector Fields}
\author{Benjam\'{i}n Grinstein}
\email{bgrinstein@ucsd.edu}
\affiliation{Department of Physics, University of California, San Diego, La Jolla, CA 92093 USA}
\author{Christopher W. Murphy}
\email{cmurphy@physics.ucsd.edu}
\affiliation{Department of Physics, University of California, San Diego, La Jolla, CA 92093 USA}
\author{Michael Trott}
\email{michael.trott@cern.ch}
\affiliation{Theory Division, Physics Department, CERN, CH-1211 Geneva 23, Switzerland}

\begin{abstract}
Electroweak precision data constraints on flavor symmetric vector fields are determined. The flavor multiplets of spin one that we examine are the complete set of fields that couple to quark bi-linears at tree level
while not initially breaking the quark global flavor symmetry group. Flavor safe vector masses proximate to, and in some cases below, the electroweak symmetry breaking scale are found to be allowed.
Many of these fields provide a flavor safe mechanism to explain the $t \, \bar{t}$ forward backward anomaly, and can simultaneously significantly raise the allowed values of the Standard Model Higgs mass 
consistent with electroweak precision data.
\end{abstract}

\maketitle
%\newpage
\section{Introduction}
The global flavor symmetry group of the Standard Model (SM) is only
broken by the Yukawa matrices.  As a result, the SM predicts a non-trivial and definite pattern in flavor changing decays and meson
mixing observables; there are no flavor changing neutral currents
(FCNC's) at tree level, and flavor changing charged currents follow
the pattern of the CKM matrix.  The predicted pattern of flavor
changing observables in the SM is consistent with what is observed in
precise measurements of ${\rm Br}(b\rightarrow s + \gamma)$, $K^0 -
\bar{K}^0$, $B^0 - \bar{B}^0$ mixing, etc.  The
lack of any apparent pattern of statistically significant deviations
in these observables\footnote{Note that some sets of measurements do
  show interesting patterns of deviations from the SM, in particular
  the measurement of the CP violating phase in $B_s$ mixing at the
  Tevatron through dimuon final states and $B \rightarrow J/\psi
  \phi$; see \cite{Ligeti:2010ia,Lenz:2010gu} for related
  discussions. The recently reported $\rm LHC_b$ measurement
  \cite{LHCb} of $B \rightarrow J/\psi \phi$ does not support a NP
  interpretation of this data, while recent updates to D\O\,
  measurements of the same decay \cite{Abazov:2011ry} are still able
  to accommodate the large phase required by the dimuon anomaly.}
places strong constraints on New Physics (NP) at the TeV scale invoked
to resolve the hierarchy problem of the electroweak symmetry breaking
(EWSB) scale, $v$, in the SM.

One way to attempt to reconcile these facts with NP that appears at
the $\sim$ TeV scale is to assume that the NP has a non-generic flavor
structure.  A popular assumption that can frequently accommodate this
tension is to assume that NP has exactly the dominant flavor breaking
pattern that has been experimentally established, {\it i.e.}, that NP
that couples to quarks has the same flavor breaking structure as in
the SM. This is known as the principle of Minimal Flavor Violation
(MFV), which states that there is a unique source of flavor symmetry
breaking \cite{Chiv:87, Hall:90, D'Am:02} at scales measurable in
flavor physics observables.  Another, closely related assumption is
that a TeV scale new physics sector is completely flavor symmetric
with respect to the quark flavor symmetry group -- $\rm G_F = U(3)_Q
\times U(3)_U \times U(3)_D$. Breaking of this flavor symmetry is not
completely absent because the quark masses already break $\rm G_F$
through Yukawa interactions
\begin{equation}
\label{Yukawa}
\mathscr{L}_Y = Y_U \, \bar{u}_R \, H^T \, i \, \sigma_2 \, Q_L - Y_D \, \bar{d}_R \, H^\dagger \, Q_L + \text{h.c.}
\end{equation}
and this breaking will appear in loop corrections to the couplings of
the NP to quark bilinears. This flavor breaking however will follow
the SM pattern and be of an MFV form.

Further flavor breaking can be MFV like or deviate from the pattern
expected in MFV.  The top quark is the only quark with an $\mathcal{O}(1)$
Yukawa coupling, $y_t$. This large coupling generally forces
phenomenologically viable $\sim v$ scale NP flavor breaking effects
$\propto y_t$ to be aligned with the SM flavor breaking.  This can be
taken to mean that flavor breaking involving the third generation
($\rm G_F \rightarrow H_F$ where $\rm H_F = U(2)_Q \times U(2)_U
\times U(2)_D \times U(1)^3$) should be aligned with the SM.  Versions
of this scenario are referred to as next to minimal flavor violation
in the literature \cite{Agashe:2005hk}.

The scalar and vector fields that can couple to the SM at dimension
four respecting SM gauge symmetry and not breaking $\rm G_F$ have
recently been classified and studied in some detail
\cite{Manohar:2006ga,Arn:09,Grin:11b}. In this paper, we examine
oblique electroweak precision data (EWPD) corrections due to the
flavor symmetric vector field multiplets of this form in much more
detail.  We seek to perform as general an analysis of EWPD as possible
on phenomenologically interesting flavor symmetric vector fields in
this paper, and consequently only consider the (large) $\rm G_F
\rightarrow H_F$ flavor breaking in the mass spectrum.  Sub-dominant
flavor breaking could be MFV-like --- where all flavor breaking comes
with SM Yukawa insertions --- or could deviate from the SM pattern of
flavor breaking. In the later case, potentially important flavor
changing observable based constraints could exist.  For oblique EWPD
constraints, only the leading flavor breaking we include is expected
to be relevant. In fact, we will show that the breaking of $\rm G_F$
is not directly related to the breaking of custodial symmetry $\rm
SU_C(2)$ to the approximation we work to, and thus we can neglect
flavor breaking in general when considering oblique EWPD constraints.
We find that vector masses $m_v \sim v$ are consistent with EWPD in
most of the allowed representations, and in certain cases in large
regions of parameter space.  The allowed vector representations
\cite{Grin:11b} are shown in Table 1.

Note that interest in flavor symmetric fields has been increased
recently due to the ability of a number of these representations to
explain the CDF and D\O\, measurements \cite{CDF:08, D0:07} of an
anomalously large $t\bar{t}$ forward backward asymmetry,
$A^{t\bar{t}}_{FB}$.  Flavor symmetric fields can in principle explain
this anomaly in a flavor safe manner \cite{Grin:11a}.  In addition,
some fields of this form can potentially explain both the $t\bar{t}$
anomaly and same sign dimuon anomaly reported in \cite{D0:10} by D\O\,
at the same time \cite{Grin:11b}.\footnote{However, it should be noted
  that recent LHC results on the $t \bar{t}$ invariant mass spectrum
  \cite{ttspec} and the measurement of time dependent $B \rightarrow
  J/\psi \phi$ have cast some doubt on the NP interpretation of
  Tevatron anomalies in $A^{t\bar{t}}_{FB}$ or the dimuon anomaly
  \cite{LHCb}.}

EWPD constrains the vector fields we study through effectively
constraining the higher dimensional operators that are present in the
theory when the NP is integrated out.  Oblique EWPD constraints can
generally be studied by using the $\rm STU$ parameters
\cite{Pesk:91,Golden:1990ig,Holdom:1990tc} which assumes that the
masses of the new states are $\sim {\rm TeV}$.  When considering
masses comparable to (or lower than) the EWSB scale, the $\rm STUWVX$
formalism of~\cite{Mak:93} which does not expand in $v^2/m_{NP}^2$ is
preferred and can be more constraining \cite{Bur:09}. The latter
formalism reduces to the $\rm STU$ parameters when large NP scales are
present and we will use the $\rm STUWVX$
formalism of~\cite{Mak:93} in this paper, as we are
interesting in studying how light the vector multiplet masses can be.
\newpage
%%%%%%%%%%%%%%%%%%%%%%%%%%%%%%%%
%
\section{Electroweak Sector of MFV Vector Lagrangians}
%
%%%%%%%%%%%%%%%%%%%%%%%%%%%%%%%%%%
%
%
Table~\ref{table vecs} lists the $\rm G_F$ symmetric representations
that couple to quark bilinears without Yukawa suppression, while
preserving SM gauge invariance.\footnote{Cases I-IV are flavor singlets and
are already discussed extensively in the literature. See
Ref.~\cite{del:10} for a recent discussion.}
In this section, we construct the gauge sector of these Lagrangians\footnote{See \cite{Grin:11b} for the Yukawa sectors of
  these models and related phenomenological constraints.} for
cases V-XI  and determine the contribution of these fields to the self energies of the
SM gauge bosons. As massive flavor symmetric vectors are effective
fields in a non-renormalizable extension of the SM, we also consider
the contribution of higher-dimension operators, which are suppressed
by the cutoff scale, $\Lambda$, of the effective
theory~\cite{Grin:91}. These operators are necessary to obtain finite
oblique corrections in some cases.  One naively expects the scaling
$\Lambda \sim 4 \, \pi \, m_v$, but the separation of these scales can
be smaller. Indeed, this might be expected to be the case due to the
relatively large number of degrees of freedom in the vector
multiplets. As a non-perturbative study is clearly beyond the scope of
this initial work, we assume that the operators suppressed by higher
powers of $\Lambda$ that we do not retain are sufficiently suppressed.
\begin{table}[b]
\centering
 \begin{tabular}{ | c | c | c | c | c | c |}
  \hline
  Case & $\rm SU(3)_c$ & $\rm SU_L(2)$ & $\rm U(1)_Y$ & $\rm SU(3)_{U_R}\times SU(3)_{D_R}\times SU(3)_{Q_L}$ & couples to \\ \hline
 I$_{\text{s,o}}$ & 1,8 & 1 & 0 & (1,1,1) & $\bar{d}_R\gamma^{\mu}d_R$\\
 II$_{\text{s,o}}$ & 1,8 & 1 & 0 & (1,1,1) & $\bar{u}_R\gamma^{\mu}u_R$\\
 III$_{\text{s,o}}$ & 1,8 & 1 & 0 & (1,1,1) & $\bar{Q}_L\gamma^{\mu}Q_L$\\
 IV$_{\text{s,o}}$ & 1,8 & 3 & 0 & (1,1,1) & $\bar{Q}_L\gamma^{\mu}Q_L$\\ \hline
 V$_{\text{s,o}}$ & 1,8 & 1 & 0 & (1,8,1) & $\bar{d}_R\gamma^{\mu}d_R$\\
 VI$_{\text{s,o}}$ & 1,8 & 1 & 0 & (8,1,1) & $\bar{u}_R\gamma^{\mu}u_R$\\
 VII$_{\text{s,o}}$ & 1,8 & 1 & -1 & ($\bar{3}$,3,1) & $\bar{d}_R\gamma^{\mu}u_R$\\
 VIII$_{\text{s,o}}$ & 1,8 & 1 & 0 & (1,1,8) & $\bar{Q}_L\gamma^{\mu}Q_L$\\
 IX$_{\text{s,o}}$ & 1,8 & 3 & 0 & (1,1,8) & $\bar{Q}_L\gamma^{\mu}Q_L$\\
 X$_{\bar{3},6}$ & $\bar{3}$,6 & 2 & -1/6 & (1,3,3) & $\bar{d}_R\gamma^{\mu}Q_R$\\
 XI$_{\bar{3},6}$ & $\bar{3}$,6 & 2 & 5/6 & (3,1,3) & $\bar{u}_R\gamma^{\mu}Q_R$\\
   \hline
  \end{tabular}
  \captionsetup{justification=raggedright}
  \caption{${\rm G_F}$ symmetric vector representations from Ref.\cite{Grin:11b}.}
  \label{table vecs}
\end{table}

%
%%%%%%%%%%%%%%%%%%%%%%%
%
\subsection{Cases V, VI, and VIII}
\label{mixingsection}
These cases are not charged under the electroweak gauge group.  The
vector fields are parameterized using Gell-Mann matrices for the
color, $\tau^A$, and the flavor, $t^B$, representations.  The
Lagrangians in the mass eigenstate basis for cases $C = \{V, VI,
VIII\}$ are
\begin{align} 
\label{L568}
\mathscr{L}_{C} &= (1+\delta_{a,o})\left(-\mathscr{L}^{\text{kin}}_{C} + \mathscr{L}^{\text{mass}}_{C}\right) + \mathscr{L}^{\text{int}}_{C}   + \mathscr{L}^{\text{Yuk}}_C + \text{h.c.,}  \\
\mathscr{L}^{\text{kin}}_{C} &= \frac{1}{2}\text{Tr}\left(V^{a;\mu\nu}V^{a; \dagger}_{\mu\nu} \right)  +\frac{\kappa}{2}\,\text{Tr}\left(V^s_{\mu\nu} \, \Delta_{C}\right)B^{\mu\nu}, \nonumber \\
\mathscr{L}^{\text{mass}}_{C} &= \left(m^2 + \lambda H^{\dagger}H\right)\text{Tr}\left(V^{a; \mu}V^{a; \dagger}_{\mu} + \zeta_1 V^a_{\mu}\Delta_{C}V^{a; \dagger}_{\mu} + \zeta_2 \Delta_{C}V^a_{\mu}\Delta_{C}V^{a; \dagger}_{\mu} +\cdots\right), \nonumber \\
\mathscr{L}^{\text{int}}_{C} &= -\beta\,\text{Tr}\left(V^s_{\mu} \, \Delta_{C}\right) H^{\dagger}D^{\mu}H, \nonumber 
\end{align}
where $a=o,s$ stands for the octet and singlet sub-cases and $V^s_{\mu
  \, \nu} = \partial_\mu V^s_\nu - \partial_\nu V^s_\mu$.  The trace
is over flavor space.  In the color octet case, a covariant derivative
is needed and the trace extends to color space.  Note that the flavor
symmetry ensures that the flavor basis of the vector fields that
couple through the operators shown is the same as the flavor basis
obtained after rotating to the quark mass field basis in $
\mathscr{L}^{\text{Yuk}}$, {\it i.e.}, that no further flavor violation is
present due to a misalignment of the flavor eigenbases.

We have included explicit insertions of Yukawa matrices in
$\mathscr{L}^{\text{mass}}_{C}$ as an illustrative example.  A series
of flavor breaking insertions are also possible on the other terms in
the Lagrangian inside the trace.  Using the formalism of MFV, the
insertions are parameterized by a series in powers of $\Delta_C =
\{Y_d^{\phantom{\dagger}}Y^{\dagger}_d, Y_u^{\phantom{\dagger}}Y^{\dagger}_u, Y^{\dagger}_uY_u^{\phantom{\dagger}}\}$ for $C =
\{V, VI, VIII\}$.  For flavor breaking insertions in the kinetic
terms, we can always re-diagonalize the fields with a finite
renormalization and we neglect the resulting mass splittings,
implicitly absorbing these splittings into the leading order mass
definition for each flavor.  The insertion of Yukawa matrices into the
mass terms causes mass splittings among the different flavors. We
retain the leading flavor breaking due to the top or bottom Yukawa
leading to the mass spectrum
\begin{equation}
m^2_{1,2,3} = m^2+\frac{\lambda}{2}v^2, \quad m^2_{4,5,6,7}  = m_1^2\left(1 + \frac{\zeta_1}{2}y^2\right),  \quad m^2_8 = m_1^2\left(1 +\frac{2 \, \zeta_1}{3}y^2 + \frac{2 \, \zeta_2}{3}y^4\right).
\end{equation}
Here we have used the conventions $\langle H\rangle = v/\sqrt{2}$ and
Tr$\,(\tau^a\tau^b)=$ Tr$\,(t^at^b)=\delta^{ab}/2$.

$\mathscr{L}^{\text{int}}_{C}$ arises only in the color singlet case,
and is suppressed by an insertion of $\Delta_C$.  This operator leads
to tree level mixing of the SM gauge bosons with the new vector
multiplet suppressed by the appropriate Yukawa matrices.  When only third
generation Yukawa matrices are retained, only the 8 flavor component mixes.
For simplicity, we only consider a single mixing with $Z, A$ and $V$,
treating $\beta, \kappa \ll 1$ rather than sum the geometric series
that results if all insertions of $\Delta_C$ are
unsuppressed.\footnote{One can always reinterpret this parameter to
  correspond to a series of insertions $\beta\Delta_C +
  \beta^{'}(\Delta_C)^2 + \cdots$.}  We treat this mixing as a
perturbation. The fields are transformed to a new field basis
$\tilde{V}_8, \tilde{B}$ with diagonalized kinetic terms in the
presence of the kinetic mixing between the $V^8$ colour singlet vector
and the $B$ field. The required transformation on the field basis is
\bea \left(\begin{array}{c}
    V_8^\alpha  \\
    B^\alpha
  \end{array} \right)
 = \left(\begin{array}{cc} 
   1  & \frac{2 \, \kappa \, y^2}{\sqrt{3}} \,  \frac{m_B^2}{m_B^2 - m_8^2} \\
    \frac{2 \, \kappa \, y^2}{\sqrt{3}} \,  \frac{m_8^2}{m_8^2 - m_B^2} & 1 \\
  \end{array} \right) \, \left(\begin{array}{c} 
   \tilde{V}_8^\alpha  \\
   \tilde{B}^\alpha
 \end{array} \right)
\eea
This transformation leaves the bare tree level mass terms of the $B$
field (after EW symmetry breaking), $m_B$, and the tree level mass of
the vector field $V_8$ unchanged.  We neglect the interaction terms in
$\mathscr{L}^{\text{Yuk}}$, that are discussed in
Ref.~\cite{Grin:11b}, assuming the direct coupling to the light quarks
is small enough that tree level vector exchanges can be neglected, and
an oblique EWPD analysis is appropriate.  Dijet constraints on the
coupling of these fields to light quarks at LHC generically constrain
this coupling to be $ \sim \mathcal{O}(0.1)$ which is consistent with
this assumption.\footnote{However, the coupling of these fields that
  involve the top quark could be far larger due to flavor splitting
  effects, allowing these fields to still explain the $A_{FB}^{t \,
    \bar{t}}$ anomaly while this EWPD analysis is appropriate; see
  \cite{Grin:11b} for a more detailed discussion.} Consistency of this
analysis also requires we neglect the tree level exchanges
$\mathcal{O}(g^2 \, \kappa^2)$ from the field redefinition of the $B$
field in the covariant derivative in the quark kinetic terms. We also
neglect $\mathcal{O}(\beta \, \kappa)$ contributions in the Lagrangian
after the field redefinitions --- these effects can be removed by a
higher order re-diagonalization of the mass and kinetic operators.
With these assumptions, the color singlet vector fields give
contributions to the self-energies of the SM gauge bosons,
\begin{align}
\label{eq568}
\Pi_{ZA}(p^2) &= 0,  \quad  \Pi_{AA}(p^2) = 0, \quad
\Pi_{ZZ}(p^2) =  \frac{|\beta|^2 \, v^2 \, m_Z^2 \, y^4}{12 \, (p^2-m_8^2)}.
\end{align}
Here $y$ is the appropriate Yukawa coupling. Only terms proportional
to $g_{\mu\nu}$ are shown.
%
%%%%%%%%%%%%%%
%
\subsection{Case VII}
Case VII is a weak singlet, but has non-zero hypercharge, $Y=-1$.
We expand the fields in terms of the color Gell-Mann matrices,
$\tau^A$.  $V^{\dagger}_{\mu} \neq V_{\mu}$ because the $V_{\mu}$
fields are in the $(\overline{3}, 3, 1)$ representation of $\rm G_F$.
The fields are flavor bi-fundamentals, $V^{\mu} =
\left(V^{\mu}\right)^i_{\,j}$, where $i$ and $j$ are the indices of the
$(\bar{3},1,1)$ and $(1,3,1)$ representations respectively.  The
Lagrangians in the mass eigenstate basis are
\begin{align}
\label{L7}
\mathscr{L}_{VII} &= (1+\delta_{a,o})\left(-\mathscr{L}^{\text{kin}}_{VII} + \mathscr{L}^{\text{mass}}_{VII}\right) + \mathscr{L}^{\text{int}}_{VII} + \mathscr{L}^{\text{c.t.}}_{VII} + \mathscr{L}^{\text{Yuk}}_{VII} +\text{h.c.},  \\
\mathscr{L}^{\text{kin}}_{VII} &= \text{Tr}\left(V^{a;\mu\nu}V^{a; \dagger}_{\mu\nu} \right) +i\,g_1\, \xi B^{\mu\nu}\, \text{Tr}\left(\left[V^a_{\mu},V^{a; \dagger}_{\nu}\right] \right), \nonumber \\
\mathscr{L}^{\text{mass}}_{VII} &= \left(m^2 + \lambda H^{\dagger}H\right)\text{Tr}\left(V^{a; \mu}V^{a; \dagger}_{\mu} + \zeta_1 V^{a; \mu}Y_uY_u^{\dagger}V^{a; \dagger}_{\mu} + \zeta_2 V^{a; \mu}Y_uY_u^{\dagger}V^{a; \dagger}_{\mu}Y_dY_d^{\dagger} +\cdots\right), \nonumber \\
\mathscr{L}^{\text{int}}_{VII} &= -\beta\,\text{Tr}\left[(V^s_{\mu}\Delta_{VII})^\dagger \right] (D^{\mu} H)^{\dagger}\tilde{H} , \nonumber 
\end{align}
where $V^s_{\mu\nu} = D_{\mu}V^s_{\nu} - D_{\nu}V^s_{\mu}$, $D_{\mu}
= \partial_{\mu} - i \,g_1\, Y \, B_{\mu}$, $\Delta_{VII} =
Y_u^{\phantom{\dagger}}Y_d^{\dagger}$, and $\tilde{H} = i \sigma_2 H^\star$ (and as before
a color gauge field term is included in the covariant derivative in
the case of color octet vector).  The mass splittings for case VII are
as follows
\begin{align}
m^2_{11,21,12,22} &= m^2+\frac{\lambda}{2}v^2, \quad \quad \quad m^2_{13,31,23,32} = m_{11}^2\left(1 +\frac{\zeta_1}{2}y^2_t\right), \\
& m^2_{33} = m_{11}^2\left(1 +\frac{\zeta_1}{2}y^2_t + \frac{\zeta_2}{2}y^2_by^2_t\right). 
\end{align}
$\mathscr{L}^{\text{c.t.}}$ contains all the terms needed to make this
theory finite, which includes higher-dimensional operators as the
theory is non-renormalizable.  See Section~\ref{sec:CT} for details on
these operators.  The vector field multiplet contributions to the
self-energies of the SM gauge bosons are,
\begin{equation}
\label{PiVII}
\begin{aligned} 
\Pi_{ZZ}(p^2) &= D(R_C)\, s_w^2\, g_1^2 \sum_{f=1}^9\, \left[f +\xi\, g +\xi^2\, h\right]\left(p^2,m_f^2\right), \\
\Pi_{ZA}(p^2) &=- D(R_C)\, s_wc_w\, g_1^2 \sum_{f=1}^9\, \left[f +\xi\, g +\xi^2\, h\right]\left(p^2,m_f^2\right),  \\
\Pi_{AA}(p^2) &= D(R_C)\, c_w^2\, g_1^2 \sum_{f=1}^9\, \left[f +\xi\, g +\xi^2\, h\right]\left(p^2,m_f^2\right),  \\
\Pi_{WW}(p^2) &= \delta_{a,s}\frac{y_t^2\, y_b^2\, |\beta|^2\, v^2\, m_W^2}{8\,(p^2-m_{33}^2)}  
\end{aligned}
\end{equation}
where $D(R_C)$ is the dimension of the color representation. Here
$s_w, c_w$ are the sine and cosine of the weak mixing angle with
convention $e=g_1/c_w$, and $g_1$ is the hypercharge coupling.  The
form factors $f, g, h$ are defined in the Appendix~\ref{appen}.

%
%%%%%%%%%%%%%%%%%%%%%%%%
%
\subsection{Case IX}
Case IX is a weak triplet that has zero hypercharge.  In addition to
being parameterized by $\tau^A$ and $t^B$ for color and flavor, the
fields are also parameterized using the Pauli matrices, $V_{\mu} =
\sigma_i V_{\mu}^i$, for the $\rm SU_L(2)$ representation.  We
suppress the color singlet, octet label on the field in this section for
clarity\footnote{The $\kappa,\beta$ operators are only for the singlet
  case as before.} while including the $\rm SU_L(2)$ index, the
Lagrangians are
\begin{align} 
\label{L9}
\mathscr{L}_{IX} &= (1+\delta_{a,o})\left(-\mathscr{L}^{\text{kin}}_{IX} + \mathscr{L}^{\text{mass}}_{IX}\right) + \mathscr{L}^{\text{int}}_{IX} + \mathscr{L}^{\text{c.t.}}_{IX} + \mathscr{L}^{\text{yuk}}_{IX} + \text{h.c.},  \\
\mathscr{L}^{\text{kin}}_{IX} &= \frac{1}{2}\,\text{Tr}\left(V_i^{\mu\nu}V^{i; \dagger}_{\mu\nu} \right)   + \frac{\kappa}{2} \text{Tr}\left(V^{i}_{\mu\nu} \, \Delta_{IX}\right)W^{\mu\nu}_i  -  g_2 \, \xi \, \epsilon_{ijk} \, W^{i; \mu\nu} \, \text{Tr}\left(V^{j}_{\mu} \, V^{k \dagger}_{\nu}\right), \nonumber \\
\mathscr{L}^{\text{mass}}_{IX} &= \left(m^2 + \lambda H^{\dagger}H\right)\text{Tr}\left(V_i^{\mu}V^{i; \dagger}_{\mu} + \zeta_1 V_i^{\mu}\Delta_{IX}V^{i \, \dagger}_{\mu} + \zeta_2 \, \Delta_{IX}V_i^{\mu}\Delta_{IX}V^{i; \dagger}_{\mu} +\cdots\right), \nonumber \\
\mathscr{L}^{\text{int}}_{IX} &= -\beta\,\text{Tr}\left(V^{i}_{\mu}\, \Delta_{IX}\right) H^{\dagger}\sigma_i D^{\mu}H, \nonumber
\end{align}
where $V^{i}_{\mu\nu} = \left(D_{\mu}V_{\nu}\right)^i -
\left(D_{\nu}V_{\mu}\right)^i$, $D^{ij}_{\mu}
= \partial_{\mu}\delta^{ij} - g_2 \, \epsilon^{ijk} \, W_{k \mu}$, and
$\Delta_{IX} = Y_u^{\dagger}Y_u^{\phantom{\dagger}}$.  The mass splittings are the same as
in Case VIII.  In this section we neglect the effects of the kinetic
mixing operator considering the case $\kappa \ll \beta, \, g_2 \, \xi,
\, g_2$.\footnote{The effect of the kinetic mixing with a nonabelian
  field has recently been studied in \cite{Chen:2009ab,Heeck:2011md}
  for example and our results can be directly extended to include
  kinetic mixing. Many of the effects of kinetic mixing can be
  absorbed into a redefinition of the remaining unknown parameters of
  this model once a diagonalization of the kinetic terms is
  undertaken.  The transformation to canonical kinetic terms is
  exactly of the form given in Section \ref{mixingsection} for each
  isospin state when the kinetic mixing is not neglected.}

The contributions to the self-energies of the SM gauge bosons are then
\begin{align} \label{piWW}
\Pi_{ZZ}(p^2) &= 2 \, D(R_C)\, c_w^2\, g_2^2 \sum_{f=1}^8\, \left[f +\xi\, g +\xi^2\, h\right]\left(p^2,m_f^2\right)  +\delta_{a,s}\frac{|\beta|^2\, y_t^4\, v^2\, m_Z^2}{12\,(p^2-m_8^2)}, \\
\Pi_{ZA}(p^2) &= 2 \, D(R_C)\, s_wc_w\, g_2^2 \sum_{f=1}^8\, \left[f +\xi\, g +\xi^2\, h\right]\left(p^2,m_f^2\right),  \nonumber \\
\Pi_{AA}(p^2) &= 2 \, D(R_C)\, s_w^2\, g_2^2 \sum_{f=1}^8\, \left[f +\xi\, g +\xi^2\, h\right]\left(p^2,m_f^2\right), \nonumber \\
\Pi_{WW}(p^2) &= 2 \, D(R_C)\, g_2^2 \sum_{f=1}^8\, \left[f +\xi\, g +\xi^2\, h\right]\left(p^2,m_f^2\right) +\delta_{a,s}\frac{|\beta|^2\, y_t^4\, v^2\, m_W^2}{3\,(p^2-m_8^2)}. \nonumber
\end{align}
%
%%%%%%%%%%%%%%
%%%%%%%%%%%%%%%%%%%%
%%%%%%%%%%%%%%%%%%%%
\subsection{Cases X and XI}
These fields are $\rm SU_L(2)$ doublets, flavor bi-fundamentals, and
are in either the color sextet or anti-triplet representations, $a= 6,
\bar{3}$.  The only difference between the two cases is they have
different hypercharges, $Y_C=\{-1/6,5/6\}$ for $C =\{X, XI\}$.  The
Lagrangians are
\begin{align}
\label{L1011}
\mathscr{L}_{C} &= -\mathscr{L}^{\text{kin}}_{C} + \mathscr{L}^{\text{mass}}_{C} + \mathscr{L}^{\text{c.t.}}_{C} + \mathscr{L}^{\text{yuk}}_{C} + \text{h.c.},  \\
\mathscr{L}^{\text{kin}}_{C} &= \frac{1}{2}V^{a;\mu\nu}V^{a; \dagger}_{\mu\nu} +i \,g_1\, \xi_1 B^{\mu\nu}\,V^a_{\mu} V^{a; \dagger}_{\nu} +i\,g_2 \, \xi_2\, V^a_{\mu}W^{\mu\nu}V^{a; \dagger}_{\nu}, \nonumber \\
\mathscr{L}^{\text{mass}}_{C} &= \left(m^2 + \lambda_1 H^{\dagger}H\right)\left(V^{a; \mu}V^{a; \dagger}_{\mu} + \zeta_1 V^{a; \mu}\Delta_CV^{a; \dagger}_{\mu} +\cdots\right) \nonumber \\
&+ \lambda_2 \left(H^{\alpha\dagger}V^{a;\mu}_{\alpha}V^{a; \beta\dagger}_{\mu}H_{\beta} +\cdots\right) + \lambda_3 \left(\tilde{H}^{\alpha\dagger}V^{a;\mu}_{\alpha}V^{a; \beta\dagger}_{\mu}\tilde{H}_{\beta} +\cdots\right), \nonumber
\end{align}
where $V^6_{\mu\nu} = D_{\mu}V^6_{\nu} - D_{\nu}V^6_{\mu}$, $D_{\mu}
= \partial_{\mu} -ig_3 \tau^A_6 A^A_{\mu} -ig_2\sigma^i W^i_{\mu}
-ig_1YB_{\mu}$, and $\Delta_C = Y_u^{\phantom{\dagger}}Y_u^{\dagger}$
in both cases.  In the last line of $\mathscr{L}^{\text{mass}}_{C}$
the weak indices, $\alpha$ and $\beta$, are explicit.  We did not
explicitly write down in the Lagrangian the flavor breaking insertions
in the additional Higgs terms that appear in the mass splittings
below.  In these cases there is a mass splitting in the electroweak
doublet
\begin{equation}
m^2_{Q+} = m^2 + (\lambda_1 + \lambda_2)\frac{v^2}{2}, \, m^2_{Q-} = m^2 +  (\lambda_1 + \lambda_3)\frac{v^2}{2}
\end{equation}
where $Q\pm = Y\pm1/2$.  Note that if there were no mass splitting
between the weak states then there would to no contribution to
$\Pi_{W^3B}$ because the interaction has the form
Tr\,$(V_{\mu}^{\dagger}\sigma^3V^{\mu})$.  In addition there is the
usual mass splitting in flavor space
\begin{equation}
m^2_{11,12,21,22}  = m_{Q\pm}^2,  \, m^2_{13,23,32,31,33}  = m_{Q\pm}^2\left(1 + \zeta y^2_t\right)
\end{equation}
The contribution of the vector fields to the self-energies of the SM
gauge bosons are written in the electroweak basis in this case to
reduce clutter,
\begin{align} \label{piWB}
\Pi_{W^3W^3}(p^2) &= D(R_C)\, \frac{g_2^2}{4} \sum_{f, L}\,  \left[f +\xi_2\, g +\xi_2^2\, h\right]\left(p^2,m_{f,L}^2\right), \\
\Pi_{W^3B}(p^2) &= D(R_C)\, \frac{g_2g_1Y}{2} \sum_{f, L}\,  (-1)^{L-1}\left[f +\frac{1}{2}(\xi_1 +\xi_2)\, g + \xi_1\xi_2\, h\right]\left(p^2,m_{f,L}^2\right), \nonumber \\
\Pi_{BB}(p^2) &= D(R_C)\, g_1^2Y^2 \sum_{f, L}\,  \left[f +\xi_1\, g +\xi_1^2\, h\right]\left(p^2,m_{f,L}^2\right), \nonumber \\
\Pi_{W^1W^1}(p^2) &=D(R_C)\, \frac{g_2^2}{2} \sum_{f}\,  \left[f +\xi_2\, g +\xi_2^2\, h\right]\left(p^2,m_{f,Q+}^2,m_{f,Q-}^2\right), \nonumber
\end{align}
where the sums over flavor $f$ and weak $L$ states run from 1 to 9 and
2 respectively.  For contributions from $\Pi_{W^3B}$, it is the
difference of weak states rather than the sum that contributes.  The
factor of $(-1)^{L-1}$ accounts for this.  For $\Pi_{WW}$, there is no
sum over weak state because both particles of the weak doublet need to
be in the loop to conserve electric charge.  Note that the mass
splitting among weak states causes $\Pi_{WW}$ to be a function of both
masses.

The $S$ parameter is negative in certain regions of parameter space for cases X and XI. For example, with $\xi_1 = \xi_2 = 0$ and $m_{\pm}^2 \gg m_Z^2$, $S \propto Y_C\, \ln\left(m_{+}^2 / m_{-}^2\right)$.

%%%%%%%%%%%%%%%%%%%
\subsection{Counterterms and Higher-Dimensional Operators}
\label{sec:CT}
%%%%%%%%%%%%%%%%%%%
The vectors are effective fields, and the Lagrangians contain an
infinite number of non-renomalizable operators.  At low scales, $p$,
compared to the cutoff scale of the effective theory, $\Lambda$, the
contribution of these terms to self-energies is suppressed by powers
of $(p/\Lambda)^n$. The ratio of scales can be set by the external
momentum $p^2/\Lambda^2$ or can be set by the ratio of other
invariants $m_v^2/\Lambda^2$ depending on the operator of interest.
Because of this suppression we neglect contributions to the
self-energies from almost all of these other operators.  The reason
why we did not neglect the contributions from all of the higher
dimensional operators is explained in what follows.

Being psuedo-obersevables, the $\rm STUVWX$ parameters must be free of
divergences and independent of the renormalization point $\mu$.
Contributions to the self-energies have two origins.  Contributions of
the first kind contain no powers of momentum in the numerator from
internal propagators.  As expected, the resulting $\rm STUVWX$
parameters are finite and $\mu$-independent. These are the only terms
that would be found in a renormalizable theory.  There is no need to
add a field renormalization term, such as $Z_B B^{\mu\nu}B_{\mu\nu}$,
to the Lagrangian because its contribution to each of these parameters
is identically zero.

The second type of contributions contain all of the other terms that
come about from at least one propagator's $p_{\mu}p_{\nu}$ piece. In
this case, contributions to the $\rm STUVWX$ parameters are divergent
and require higher-dimensional operators as counter terms.
$\mathscr{L}^{\text{c.t.}}_{C}$ contains all the counterterms
necessary to absorb the divergences of the self-energies from the
dimension-4 operators.  Not all of the operators are needed as
counterterms in each case. With this understanding, the counterterms are
\begin{align}
\label{eqn:CT}
\mathscr{L}^{\text{c.t.}}_{C} &= \frac{Z_1}{\Lambda^2}\partial_{\rho}B_{\mu\nu}\partial^{\rho}B^{\mu\nu} +\frac{Z_2}{\Lambda^4}\partial_{\tau}\partial_{\rho}B_{\mu\nu}\partial^{\tau}\partial^{\rho}B^{\mu\nu} +\frac{Z_3}{\Lambda^2}\left|H^{\dagger}D_{\mu}H\right|^2 
 \\
& +\frac{Z_4}{\Lambda^2}\text{Tr}\left(D_{\rho}W_{\mu\nu}D^{\rho}W^{\mu\nu}\right) +\frac{Z_5}{\Lambda^4}\text{Tr}\left(D_{\tau}D_{\rho}W_{\mu\nu}D^{\tau}D^{\rho}W^{\mu\nu}\right) \nonumber \\
& +\frac{Z_6}{\Lambda^2}H^{\dagger}W_{\mu\nu}H B^{\mu\nu} +\frac{Z_7}{\Lambda^4}H^{\dagger}D_{\rho}W_{\mu\nu}H\partial^{\rho}B^{\mu\nu} +\frac{Z_8}{\Lambda^6}H^{\dagger}D_{\tau}D_{\rho}W_{\mu\nu}H\partial^{\tau}\partial^{\rho}B^{\mu\nu}  \nonumber
\end{align}
The divergences come in the form $2/\varepsilon -\gamma
+\ln{(4\pi\mu^2/m^2)}$ (using dimensional-regularization in $D = 4 -
\varepsilon$ dimensions) and we use the operators to cancel the
divergences and $\mu$ dependence. In practice, we take $\mu = 1$ TeV
for numerical evaluations and include in our fit a finite contribution
from one higher-dimensional operator to absorb the $\mu$-dependence
when a divergence is canceled by a higher dimensional operator.
%%%%%%%%%%%%%%%%%%%%%%%%%%%%%%%%%%%%%%%%%%%%
%
%%%%%%%%%%%%%
\section{Fit to Electroweak Precision Data}
\label{ewpd}
%
%%%%%%%%%%%%%%%%%
%
Electroweak precision data provides strong constraints on the MFV
vectors under consideration.  A convenient subset of the full set of
corrections are the oblique corrections.  Purely NP contributions to
the self-energies of the electroweak gauge bosons can be written in
the following form $\Pi^{\mu\nu}_{ab}(k) = \Pi(k^2)_{ab}g^{\mu\nu} +
f(k^2)k^{\mu}k^{\nu}$ where $ab = \{W^{+}W^{-}, ZZ, AA, ZA\}$.  When
the masses of the new states are heavy compared to the EWSB scale, the
vacuum polarizations can be expanded in momentum $\Pi_{ab}(q^2)
\approx \Pi_{ab}(0) +q^2\Pi^{'}_{ab}(0)$ and the $\rm STU$ parameters
can be used. However, when one wishes to consider masses comparable to
(or smaller than) the EWSB scale, this expansion in momentum becomes
invalid.  In that case, the $\rm STUWVX$ parameters (defined in
\cite{Mak:93}), which do not expand in $v^2/m_{NP}^2$, can be used and
can be more constraining.  The theoretical predictions of EWPD of the
2008 PDG~\cite{PDG:08} was used to perform a modern fit to the $\rm
STUVWX$ parameters in Ref.~\cite{Bur:09}.  The results of the fit are
given in Table~\ref{fit}.
\begin{table}[b]
 \centering
 \begin{tabular}{ | c | c | c | }
  \hline
  Oblique & $y \pm \sigma$ \\ \hline
S & $0.07 \pm 0.41$ \\
T & $-0.40 \pm 0.28$ \\
U & $0.65 \pm 0.33$ \\
V & $0.43 \pm 0.29$ \\
W & $3.0 \pm 2.5$ \\
X & $-0.17 \pm 0.15$ \\
   \hline
  \end{tabular}
 \captionsetup{justification=raggedright}
  \caption{EWPD fit to STUVWX results of Ref.~\cite{Bur:09}. $y$ is the best-fit value, and $\sigma$ is the square root of the diagonal element of the determined covariance matrix.}
  \label{fit}
\end{table}
The correlation coefficient matrix of the fit is given by
\begin{equation}
\rho = 
\begin{pmatrix}
1 & 0.60 & 0.38 & -0.57 & 0 & -0.86 \\
0.60 & 1 & -0.49 & -0.95 & 0 & -0.13 \\
0.38 & -0.49 & 1 & 0.46 & -0.01 & -0.76 \\
-0.57 & -0.95 & 0.46 & 1 & 0 & 0.13 \\
0 & 0 & -0.01 & 0 & 1 & 0 \\
-0.86 & -0.13 & -0.76 & 0.13 & 0 & 1
\end{pmatrix}.
\end{equation}

The least-squared estimators $\boldsymbol{\hat{\theta}}$ for the set of parameters $\boldsymbol{\theta}$ of a given model, is determined by the minimum of  
\begin{equation}
\label{chi}
\chi^2(\boldsymbol{\theta}) = \left(\mathbf{y}-\mathbf{F}(\boldsymbol{\theta})\right)^TV^{-1}\left(\mathbf{y}-\mathbf{F}(\boldsymbol{\theta})\right) 
\end{equation}
where $\mathbf{y}$ is a vector of the best-fit values of the $\rm STUVWX$ parameters, $V^{-1}$ is the inverse of the covariance matrix $V_{ij}=\sigma_i\rho_{ij}\sigma_j$, and $\mathbf{F}(\boldsymbol{\theta})$ is the corresponding vector of predicted values of the model.  We determine the 1- and 2-$\sigma$ confidence regions of allowed parameter space by requiring that the contribution of NP to $\chi^2(\boldsymbol{\theta})$ satisfy
\begin{equation}
\chi^2(\boldsymbol{\theta}) \leq \chi^2_{min} + \Delta\chi^2
\end{equation}
where $\Delta\chi^2$ corresponds to the probability that the confidence region of parameter space determined with the cumulative distribution function and contains the best fit value of the parameters.  
%%%%%%%%%%%%%%%%%%%%%%%%%%%%%%%%%%%%%%%
%%%%%%%%%%%%
\section{Results} 
\label{figs} 
In this section, constraints on flavor symmetric vectors are discussed
on a case-by-case basis.  Mass splittings among flavor states are
generally assumed to be small and not relevant for these constraints.
This follows from the fact that there are no loops with different
flavor multiplet species to the approximation we work to, so the
breaking of $\rm G_F$ is not linked to the breaking of custodial
symmetry $\rm SU_C(2)$.  Note however that in certain cases, such as
the constraint plots for Section A, what is plotted is the constraint
space for $\beta$ and $m_8$. The other vectors in these multiplets are
split in general from the mass $m_8$ with a splitting of
$\mathcal{O}(\zeta_{1,2})$ and the appropriate Yukawa suppression.

\subsection{Cases V, VI, and VIII}
In all of these color octet cases, EWPD does not place limits on the
parameters of the model in the approximation that we are working to
--- the relevant operator is forbidden.  However, the absence of
vector pair-production at LEP implies a kinematic bound for color
singlet and octet vectors of at least $m_V \gsim 105$ GeV.  When the
direct coupling of the vectors to quarks is $\mathcal{O}(1)$, the experimental
bound become stronger, $m_V \gsim 150$ GeV from anomalous multi-jet
events at LEP~\cite{Grin:11b}. We also include the latter bound in the
figure.  Dedicated collider searches can significantly raise these
mass bounds.

\begin{figure}
  \centering
   \includegraphics[width=1\textwidth]{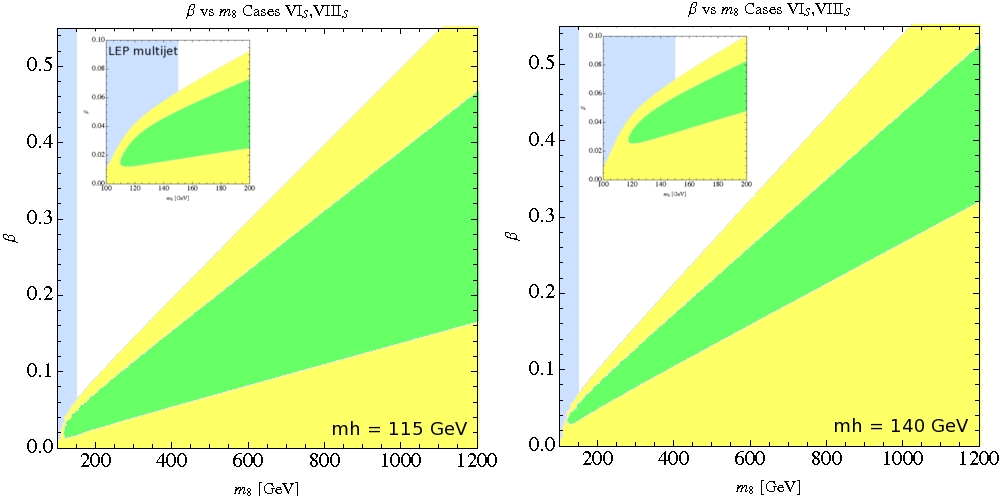}
    \captionsetup{justification=raggedright}
     \vspace{-1cm}
 \caption{In all of the plots, the green and yellow regions represent
   regions of parameter space that are allowed at 1- and 2-$\sigma$
   respectively. Also shown is the approximate bound from LEP due to
   multijet final states when the coupling to light quarks is not
   neglected. Note that these bounds are complementary in that the
   oblique analysis fails in this case.}
\label{fig1}
\end{figure}
The bounds from EWPD on case V$_{\text{s}}$ are particularly weak as
$y_b \ll 1$ and the other parameters in this model are unconstrained.
In theories such as the large $\tan \beta$ limit of the MSSM where
$y_b$ becomes $\mathcal{O}(1)$, the constraints are similar to cases
VI$_{\text{s}}$ and VIII$_{\text{s}}$. In Fig.~\ref{fig1} the Higgs mass is
fixed to $m_h = 115 (140) \, {\rm GeV}$ in the left (right) figure
with its one loop contribution to the EW parameters floated from a
reference value of $\hat{m}_h = 96 \, {\rm GeV}$ in the fit.

\begin{figure}
  \centering
   \includegraphics[width=1\textwidth]{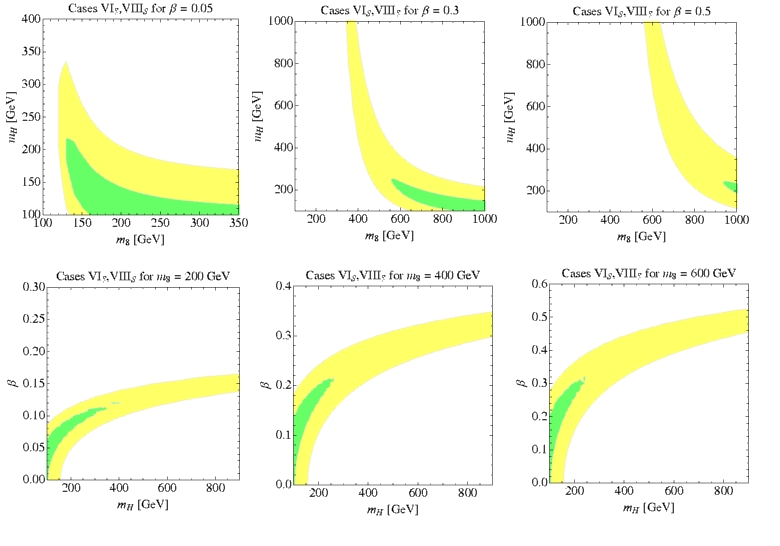}
    \captionsetup{justification=raggedright}
    \vspace{-1cm}
 \caption{Allowed confidence regions at 1- and 2-$\sigma$ when the
   Higgs mass is raised through a $\rm SU_C(2)$ violating operator -
   due to coupling to a flavor symmetric vector multiplet.}
\label{fig2}
\end{figure}

Operators such as $\text{Tr}\left(V^s_{\mu} \, \Delta_{C}\right)
H^{\dagger}D^{\mu}H$ lead to a violation of $\rm SU_C(2)$ and can act
to raise the Higgs mass by giving a positive contribution to the $T$
parameter.  We illustrate the effect on the best fit value of the
Higgs mass in Fig.~\ref{fig2}. The entire light mass region of the Higgs in
the SM may be excluded in the near future by CMS and/or ATLAS or a
combination of the experimental data sets.  Simple (flavor safe)
mechanisms to raise the Higgs mass in the EWPD as demonstrated here
would then be of greater interest. This mechanism also exists in
several of the remaining cases when an operator of this form is
allowed.  We will generally fix the Higgs mass in what follows to
reduce the parameter space.

\subsection{Case VII}
\begin{figure}
  \centering
   \includegraphics[width=1.05\textwidth]{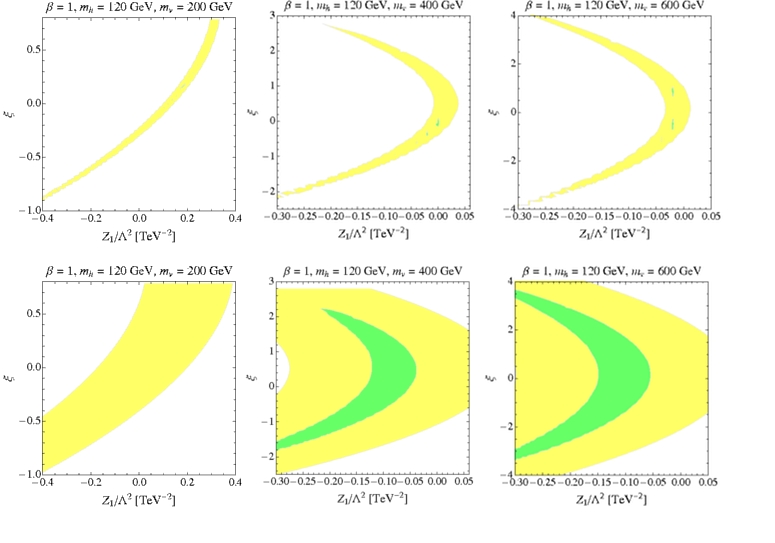}
    \captionsetup{justification=raggedright}
    \vspace{-1cm}
 \caption{Space of allowed $\xi$ and $Z_1/\Lambda^2$ for cases $\rm
   VII_o$ (top) and $\rm VII_s$ (bottom). Here we have set
   $\zeta_{1,2} = 0$, neglecting flavour breaking. The regions shown
   are weakly dependent on variations in $\beta$, we have fixed $\beta
   = 1$ and $m_h = 120 \, {\rm GeV}$. Left to right the masses are
   $m_v = 200, 400, 600 \, {\rm GeV}$.}
\label{fig:7sa}
\end{figure}

\begin{figure}
\centering
   \includegraphics[width=1\textwidth]{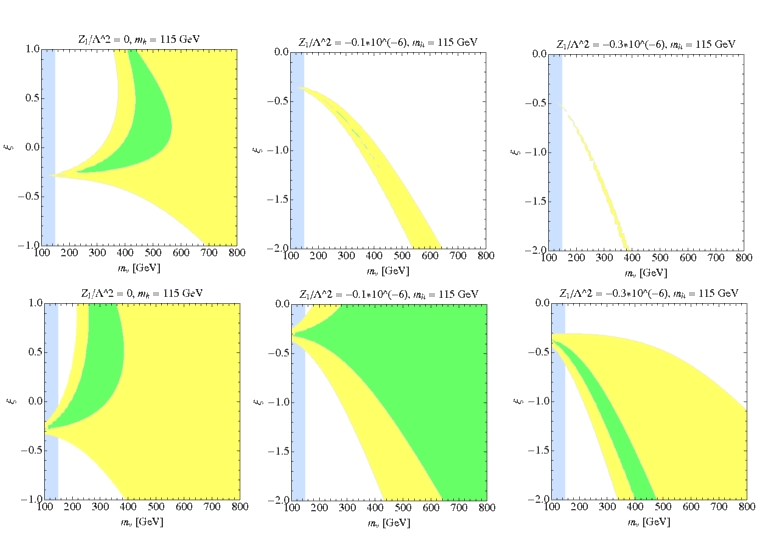}
    \captionsetup{justification=raggedright}
    \vspace{-1cm}
 \caption{Space of allowed $\xi$ and $m_V$ for cases $\rm VII_o$ (top)
   and $\rm VII_s$ (bottom). Here we neglecting flavor breaking and
   $\beta$ as the allowed parameter space is weakly dependent on these
   parameters. We have set $m_h = 115 \, {\rm GeV}$. Left to right the
   counterterm values are $Z_1/\Lambda^2 = (0,-0.1,-0.3) \times \,
   {\rm TeV}^{-2}$.}
\label{fig:7sb}
\end{figure}

Eliminating the $\mu$-dependence from $\rm STUVWX$ determines a
relationship between $\xi$ and $Z_1/\Lambda^2$ as a function of the
other (allowed) parameters in the model.  For numerical purposes, we
ignore the dimension-8 operator with coefficient $Z_2$.  Fig.~\ref{fig:7sa} shows
the relationships between $\xi$ and $Z_1/\Lambda^2$ for various
masses.
The allowed regions in the multidimensional parameter space has a
nontrivial dependence on the various parameters, as is further
illustrated in Fig.~\ref{fig:7sb}.  Generically the parameter space for $\rm
VII_s$ is less constrained than $\rm VII_o$. Both cases require a
strong correlation between the finite part of the counter term and the
remaining parameters to not be ruled out.

\subsection{Case IX}
\begin{figure}
\centering
 \includegraphics[width=1.05\textwidth]{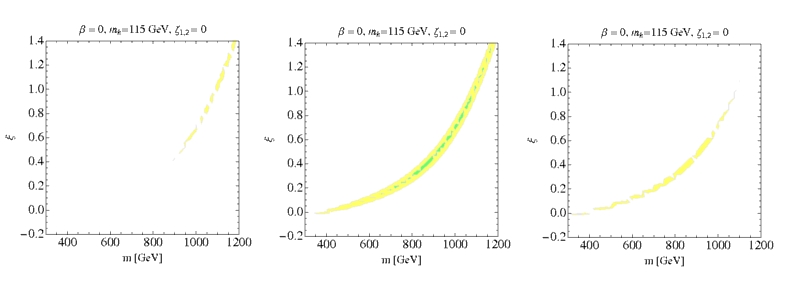}
    \captionsetup{justification=raggedright}
    \vspace{-1cm}
 \caption{Space of allowed $\xi$ and $m$ for case $\rm IX_s$. Here we
   have set $\beta = 0$ and used $m_h = 115 \, {\rm GeV}$. Left to
   right the counterterm is $Z_4/\Lambda^2 =
   (-3,0,3) \times 10^{-2} \, {\rm TeV^{-2}}$.}
 \label{fig5}
 \end{figure}

The operator proportional to $\beta$ again leads to a relaxation of
the Higgs mass bounds as in Cases $\rm VI_s$, $\rm VIII_s$. We set
$\beta = 0$ and $m_h = 115 \,{\rm GeV}$ in what follows and examine
the remaining parameter space. This case has the strongest constraints
from oblique EWPD. A strong correlation is required between $\xi$, $m$
and $Z_4/\Lambda^2$ for the allowed parameter space.  This is
illustrated in Fig.~\ref{fig5} for $\rm IX_s$. We do not find viable
parameter space for $\rm IX_o$.

\subsection{Cases X and XI}
\begin{figure}
  \centering
   \includegraphics[width=1.1\textwidth]{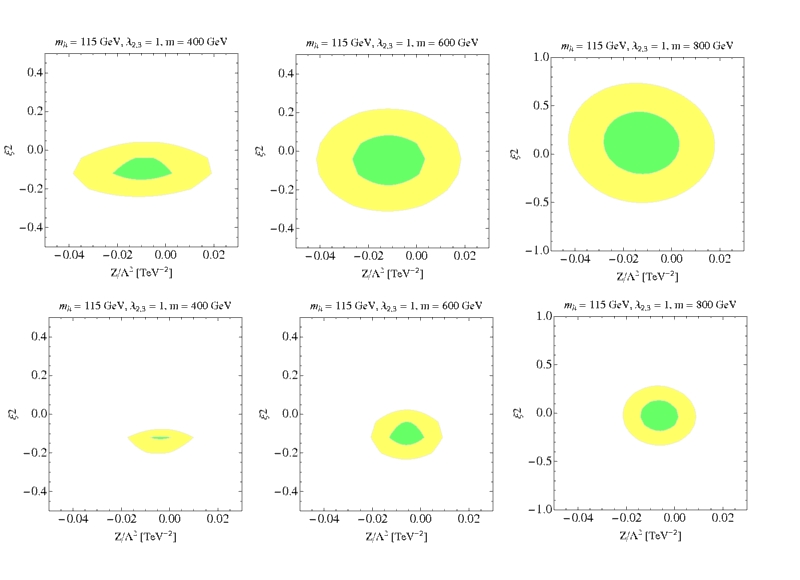}
   \captionsetup{justification=raggedright}
   \vspace{-1cm}
 \caption{Space of allowed $\xi_2$ and $Z/\Lambda^2$ for cases $\rm
   X_{\bar{3}}$ (top) and $\rm X_{6}$ (bottom). Here we neglect
   flavour breaking and $\xi_1$.  Breaking of the mass degeneracy of
   the $\rm SU_L(2)$ states is included, again the effect on the
   parameter space is negligible.  Left to right the masses are $m_v =
   400,600,800 \, {\rm GeV}$.}
\label{fig:10a}
\end{figure}

\begin{figure}
  \centering
   \includegraphics[width=1.05\textwidth]{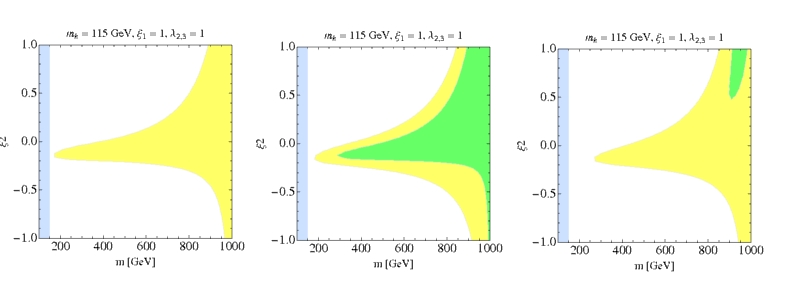}
    \captionsetup{justification=raggedright}
    \vspace{-1cm}
    \caption{Space of allowed $\xi_2$ and $m$ for  $\rm
      X_{\bar{3}}$. Flavour breaking is neglected and we set
      $\xi_1 = 1$ and $\lambda_{2,3} = 1$ - the dependence on these
      parameters is negligible. We have also set $m_h = 115 \,
      {\rm GeV}$.  Left to right $Z/\Lambda^2 = (0.01,-0.01,-0.03) \,
      {\rm TeV^{-2}}$. For case $\rm X_{6}$ the parameter space with
      $Z/\Lambda^2 = -0.01\, {\rm TeV^{-2}}$ is  similar, while 
      there is no allowed parameter space for the other values of
      $Z/\Lambda^2$.}
\label{fig:10b}
 \centering
   \includegraphics[width=1.05\textwidth]{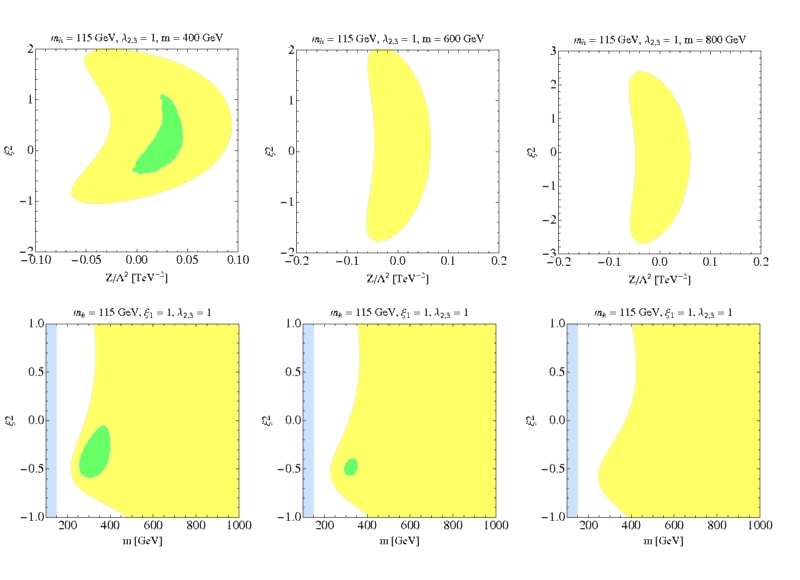}
    \captionsetup{justification=raggedright}
    \vspace{-1cm}
    \caption{Space of allowed $\xi_2$ and $Z/\Lambda^2$ for case $\rm
      XI_{\bar{3}}$ (top), left to right the masses are $m_v =
      400,600,800 \, {\rm GeV}$, other parameters same as above,  $Z/\Lambda^2$ same as above left to right for $\rm
      XI_{\bar{3}}$ (bottom row).}
 \label{fig:11a}
   \end{figure}

There is no operator proportional to $H^{\dagger}D^{\mu}H$ in these
cases that directly violates $\rm SU_C(2)$ and allows the Higgs mass
to be raised.  In cases $\rm X_{6, \bar{3}}$ the allowed parameter
space dependence on the parameter $\xi_1$ is trivial, not showing a
significant correlation with $Z/\Lambda^2$, $\xi_2$ or $m_V$.  The
correlation between $\xi_2$ and $Z/\Lambda^2$ in the allowed parameter
space is shown in Fig.~\ref{fig:10a}. The allowed masses in a joint fit with
fixed $m_h = 115 \, {\rm GeV}$ is shown in Fig.~\ref{fig:10b} for model $\rm
X_{\bar{3}}$. For comparison the required correlations between the
parameters for case $\rm XI_{\bar{3}}$ are shown in Fig.~\ref{fig:11a}.
%%%%%%%%%%%%%%%%%%%%%%%%%%%%%%555
\section{Conclusions}
%%%%%%%%%%%%%%%%%%%%%%%%%%%%%%3
In this paper we have examined the constraints that oblique EWPD
places on flavor symmetric vector fields. We have examined these
constraints on the vector field multiplets that transform under the
flavor group and couple to quark bi-linears at tree level, while not
initially breaking the quark global flavor symmetry group. These
extensions to the SM are treated as effective fields in fitting to
oblique EWPD, including appropriate counterterms to make the EWPD
pseudo-observables finite and renormalization scale independent when
required.  We have found that large regions of parameter space exist
where a joint fit to these fields with the Higgs allows a good fit,
while the masses of the vector multiplets are $\sim v$.  Vectors of
this form can act to significantly relax the mass bound on the Higgs
in a flavor safe manner, as we have demonstrated in detail for Cases
$\rm VI_s$, and $\rm VIII_s$.  Flavor safe mechanisms to raise the
Higgs mass bound may be of greater interest if the entire light Higgs
mass parameter space is excluded experimentally in the near future.

Conversely, it is interesting to note that large regions of parameter
space exist in the models where the Higgs mass is in the light mass
region with $m_h \sim 115 \, {\rm GeV}$, joint fits to EWPD are
improved over the SM alone, and field content is allowed that could
possibly explain the $A_{FB}^{t \, \bar{t}}$ anomaly. Vector fields of
the form we have considered are relatively unconstrained by indirect
searches in flavor physics due to their flavor symmetry, and have been
shown to be consistent with oblique EWPD constraints. 
Further dedicated studies of the constraints on these flavour multiplets from
non oblique precision EW observables, such as  $R_b$, may provide stronger constraints on the allowed mass scales.
Dedicated direct collider studies of flavor safe vector fields also have to potential to
raise the mass bounds on these models, or discover models of this form
at LHC.

%%%%%%%%%
\begin{acknowledgements}
  The work of B.G. and C.M.  was supported in part by the US Department
  of Energy under contract DOE-FG03-97ER40546.  This material is based upon work supported in part by the National Science Foundation under Grant No. 1066293 and the hospitality of the Aspen Center for Physics.
\end{acknowledgements}

\appendix
\section{Self-Energy Form Factors} \label{appen}
Throughout this work the following form factors appear:
\begin{align*}
f(p^2,m^2,M^2) &= \frac{1}{576\pi^2m^2M^2}\left[ 3\left(2p^2\left(m^2 + M^2\right) - m^4 - 14m^2M^2 - 9M^4\right)A_0\left(M^2\right) \right. \\
&\left. + 3\left(2p^2\left(m^2 + M^2\right) - 9m^4 - 14m^2M^2 - M^4\right)A_0\left(m^2\right) \right. \\
&\left. + 3\left(24m^2M^2\left(m^2 + M^2\right) + p^2\left(3m^4 - 2m^2M^2 +3M^4\right)  \right.\right. \\
&\left.\left. \,\,\,\,\,\, - 2p^4\left(m^2 + M^2\right)\right)B_0\left(p^2, m^2, M^2\right) \right. \\
&\left. + 2\left(3\left(m^6 + 11m^4M^2 + 11m^2M^4 + M^6\right) -4p^2\left(m^4 + m^2M^2 + M^4\right)  \right.\right. \\
&\left.\left.  \,\,\,\,\,\, + p^4\left(m^2 + M^2\right) \right) \right. \\
&\left. + 3\frac{\left(m^2 - M^2\right)^2}{p^2}\left(m^4 + 10m^2M^2 + M^4\right)\left(B_0\left(0, m^2, M^2\right)  - B_0\left(p^2, m^2, M^2\right)\right)\right], \\
g(p^2,m^2,M^2) &= \frac{m^2+M^2}{32\pi^2m^2M^2}\left[\left(2p^2\left(m^2 + M^2\right) - p^4 - \left(m^2 - M^2\right)^2\right)B_0\left(p^2,m^2,M^2\right) \right. \\
&\left. +\left(m^2 - M^2 + p^2\right)A_0\left(m^2\right) + \left(M^2 - m^2 + p^2\right)A_0\left(M^2\right)\right], \\
h(p^2,m^2,M^2) &= \frac{1}{576\pi^2m^2M^2}\left[6\left(m^2 - M^2\right)^2\left(m^2 + M^2\right)\left(B_0\left(0, m^2, M^2\right) - B_0\left(p^2, m^2, M^2\right)\right) \right. \\
&\left. +3p^2\left(\left(3m^4 + 10m^2M^2 + 3M^4\right)B_0\left(p^2, m^2, M^2\right) + \left(M^2 - 9m^2\right)A_0\left(m^2\right) \right.\right. \\
&\left.\left.  \,\,\,\,\,\, + \left(m^2 - 9M^2\right)A_0\left(M^2\right) - 2\left(m^2 + M^2\right)^2\right) \right. \\
&\left. +p^4\left(3\left(A_0\left(m^2\right) + A_0\left(M^2\right)\right) +  8\left(m^2 + M^2\right)\right)  - p^6\left(3B_0\left(p^2, m^2, M^2\right) + 2\right)\right].
\end{align*}
The above expressions simplify when the particles in the loop have the
same mass.  We define $f(p^2,m^2) \equiv f(p^2,m^2,m^2)$ as the form
factor when the masses are equal.

\begin{align*}
f(p^2,m^2) = \frac{1}{144\pi^2m^2}&\left[3\left(12m^4 + m^2p^2 - p^4\right)B_0\left(p^2, m^2, m^2\right) \right. \\
&\left. +6\left(p^2 - 6m^2\right)A_0\left(m^2\right) + 36m^4 - 6m^2p^2 + p^4\right], \\
g(p^2,m^2) = \frac{p^2}{16\pi^2m^2}&\left[\left(4m^2 - p^2\right)B_0\left(p^2,m^2,m^2\right) +2A_0\left(m^2\right)\right], \\
h(p^2,m^2) = \frac{p^2}{576\pi^2m^4}&\left[3\left(16m^4 - p^4\right)B_0(p^2,m^2,m^2) + 6\left(p^2 - 8m^2\right)A_0(m^2)\right. \\
&\left. - 2\left(12m^4 - 8m^2p^2 +p^4\right)\right].
\end{align*}
Notice that when the masses in the above form factors are equal, the
form factor vanishes at zero-momentum. This ensures that gauge
invariance, which requires $\Pi_{AA}(0)= \Pi_{ZA}(0) = 0 $, is
satisfied. It also ensures that the STUVWX parameters have the correct
limiting forms. Consider $S$ in model VII, for example: since the
Taylor expansion of any form factor about $p^2=0$ starts at one
derivative one sees from Eqs.~\ref{PiVII} that the one
derivative contribution to the $S$ parameter vanishes. This is
precisely as expected from the Peskin-Takeuchi definition of S,
proportional to $\Pi'_{3B}(0)$: in model VII the vector boson couples
to $B$ but not to $W^3$.

The one-loop form factors have been written in terms of
Passarino-Veltman functions
\begin{align*}
A_0\left(m^2\right) &= 16\pi^2\mu^{4-D}\int \! \frac{d^Dq}{i(2\pi)^D} \, \frac{1}{q^2 - m^2 + i\epsilon}, \\
B_0\left(p^2,m^2,M^2\right) &= 16\pi^2\mu^{4-D}\int \! \frac{d^Dq}{i(2\pi)^D} \,  \frac{1}{\left[q^2 - m^2 + i\epsilon\right]\left[(q + p)^2 - M^2 + i\epsilon\right]}. 
\end{align*}

\bibliography{MFVv6}

\begin{thebibliography}{26}
\expandafter\ifx\csname natexlab\endcsname\relax\def\natexlab#1{#1}\fi
\expandafter\ifx\csname bibnamefont\endcsname\relax
  \def\bibnamefont#1{#1}\fi
\expandafter\ifx\csname bibfnamefont\endcsname\relax
  \def\bibfnamefont#1{#1}\fi
\expandafter\ifx\csname citenamefont\endcsname\relax
  \def\citenamefont#1{#1}\fi
\expandafter\ifx\csname url\endcsname\relax
  \def\url#1{\texttt{#1}}\fi
\expandafter\ifx\csname urlprefix\endcsname\relax\def\urlprefix{URL }\fi
\providecommand{\bibinfo}[2]{#2}
\providecommand{\eprint}[2][]{\url{#2}}

\bibitem[{\citenamefont{Ligeti et~al.}(2010)\citenamefont{Ligeti, Papucci,
  Perez et~al.}}]{Ligeti:2010ia}
\bibinfo{author}{\bibfnamefont{Z.}~\bibnamefont{Ligeti}},
  \bibinfo{author}{\bibfnamefont{M.}~\bibnamefont{Papucci}},
  \bibinfo{author}{\bibfnamefont{G.}~\bibnamefont{Perez}},
  \bibnamefont{et~al.}, \bibinfo{journal}{Phys. Rev. Lett.}
  \textbf{\bibinfo{volume}{105}}, \bibinfo{pages}{131601}
  (\bibinfo{year}{2010}), \eprint{1006.0432}.

\bibitem[{\citenamefont{Lenz et~al.}(2011)}]{Lenz:2010gu}
\bibinfo{author}{\bibfnamefont{A.}~\bibnamefont{Lenz}} \bibnamefont{et~al.},
  \bibinfo{journal}{Phys.Rev.} \textbf{\bibinfo{volume}{D83}},
  \bibinfo{pages}{036004} (\bibinfo{year}{2011}), \eprint{1008.1593}.

\bibitem[{\citenamefont{Raven}(2011)}]{LHCb}
\bibinfo{author}{\bibfnamefont{G.}~\bibnamefont{Raven}},
  \bibinfo{journal}{Lepton-Photon, Aug 27}  (\bibinfo{year}{2011}).

\bibitem[{\citenamefont{Abazov et~al.}(2011)}]{Abazov:2011ry}
\bibinfo{author}{\bibfnamefont{V.~M.} \bibnamefont{Abazov}}
  \bibnamefont{et~al.} (\bibinfo{collaboration}{D0 Collaboration})
  (\bibinfo{year}{2011}), \eprint{1109.3166}.

\bibitem[{\citenamefont{Chivukula and Georgi}(1987)}]{Chiv:87}
\bibinfo{author}{\bibfnamefont{R.}~\bibnamefont{Chivukula}} \bibnamefont{and}
  \bibinfo{author}{\bibfnamefont{H.}~\bibnamefont{Georgi}},
  \bibinfo{journal}{Phys.Lett.} \textbf{\bibinfo{volume}{B188}},
  \bibinfo{pages}{99} (\bibinfo{year}{1987}).

\bibitem[{\citenamefont{Hall and Randall}(1990)}]{Hall:90}
\bibinfo{author}{\bibfnamefont{L.}~\bibnamefont{Hall}} \bibnamefont{and}
  \bibinfo{author}{\bibfnamefont{L.}~\bibnamefont{Randall}},
  \bibinfo{journal}{Phys.Rev.Lett.} \textbf{\bibinfo{volume}{65}},
  \bibinfo{pages}{2939} (\bibinfo{year}{1990}).

\bibitem[{\citenamefont{D'Ambrosio et~al.}(2002)}]{D'Am:02}
\bibinfo{author}{\bibfnamefont{G.}~\bibnamefont{D'Ambrosio}}
  \bibnamefont{et~al.}, \bibinfo{journal}{Nucl.Phys.}
  \textbf{\bibinfo{volume}{B645}}, \bibinfo{pages}{155} (\bibinfo{year}{2002}),
  \eprint{hep-ph/0207036}.

\bibitem[{\citenamefont{Agashe et~al.}(2005)\citenamefont{Agashe, Papucci,
  Perez, and Pirjol}}]{Agashe:2005hk}
\bibinfo{author}{\bibfnamefont{K.}~\bibnamefont{Agashe}},
  \bibinfo{author}{\bibfnamefont{M.}~\bibnamefont{Papucci}},
  \bibinfo{author}{\bibfnamefont{G.}~\bibnamefont{Perez}}, \bibnamefont{and}
  \bibinfo{author}{\bibfnamefont{D.}~\bibnamefont{Pirjol}}
  (\bibinfo{year}{2005}), \eprint{hep-ph/0509117}.

\bibitem[{\citenamefont{Manohar and Wise}(2006)}]{Manohar:2006ga}
\bibinfo{author}{\bibfnamefont{A.~V.} \bibnamefont{Manohar}} \bibnamefont{and}
  \bibinfo{author}{\bibfnamefont{M.~B.} \bibnamefont{Wise}},
  \bibinfo{journal}{Phys.Rev.} \textbf{\bibinfo{volume}{D74}},
  \bibinfo{pages}{035009} (\bibinfo{year}{2006}), \eprint{hep-ph/0606172}.

\bibitem[{\citenamefont{Arnold et~al.}(2010)\citenamefont{Arnold, Pospelov,
  Trott, and Wise}}]{Arn:09}
\bibinfo{author}{\bibfnamefont{J.~M.} \bibnamefont{Arnold}},
  \bibinfo{author}{\bibfnamefont{M.}~\bibnamefont{Pospelov}},
  \bibinfo{author}{\bibfnamefont{M.}~\bibnamefont{Trott}}, \bibnamefont{and}
  \bibinfo{author}{\bibfnamefont{M.~B.} \bibnamefont{Wise}},
  \bibinfo{journal}{JHEP} \textbf{\bibinfo{volume}{1001}}, \bibinfo{pages}{073}
  (\bibinfo{year}{2010}), \eprint{0911.2225}.

\bibitem[{\citenamefont{Grinstein
  et~al.}(2011{\natexlab{a}})\citenamefont{Grinstein, Kagan, Trott, and
  Zupan}}]{Grin:11b}
\bibinfo{author}{\bibfnamefont{B.}~\bibnamefont{Grinstein}},
  \bibinfo{author}{\bibfnamefont{A.~L.} \bibnamefont{Kagan}},
  \bibinfo{author}{\bibfnamefont{M.}~\bibnamefont{Trott}}, \bibnamefont{and}
  \bibinfo{author}{\bibfnamefont{J.}~\bibnamefont{Zupan}}
  (\bibinfo{year}{2011}{\natexlab{a}}), \eprint{1108.4027}.

\bibitem[{\citenamefont{Aaltonen et~al.}(2008)}]{CDF:08}
\bibinfo{author}{\bibfnamefont{T.}~\bibnamefont{Aaltonen}} \bibnamefont{et~al.}
  (\bibinfo{collaboration}{CDF}), \bibinfo{journal}{Phys.Rev.Lett.}
  \textbf{\bibinfo{volume}{101}}, \bibinfo{pages}{202001}
  (\bibinfo{year}{2008}), \eprint{0806.2472}.

\bibitem[{\citenamefont{Abazov et~al.}(2008)}]{D0:07}
\bibinfo{author}{\bibfnamefont{V.}~\bibnamefont{Abazov}} \bibnamefont{et~al.}
  (\bibinfo{collaboration}{D0}), \bibinfo{journal}{Phys.Rev.Lett.}
  \textbf{\bibinfo{volume}{100}}, \bibinfo{pages}{142002}
  (\bibinfo{year}{2008}), \eprint{0712.0851}.

\bibitem[{\citenamefont{Grinstein
  et~al.}(2011{\natexlab{b}})\citenamefont{Grinstein, Kagan, Trott
  et~al.}}]{Grin:11a}
\bibinfo{author}{\bibfnamefont{B.}~\bibnamefont{Grinstein}},
  \bibinfo{author}{\bibfnamefont{A.~L.} \bibnamefont{Kagan}},
  \bibinfo{author}{\bibfnamefont{M.}~\bibnamefont{Trott}},
  \bibnamefont{et~al.}, \bibinfo{journal}{Phys.Rev.Lett.}
  \textbf{\bibinfo{volume}{107}}, \bibinfo{pages}{012002}
  (\bibinfo{year}{2011}{\natexlab{b}}), \eprint{1102.3374}.

\bibitem[{\citenamefont{Abazov et~al.}(2010)}]{D0:10}
\bibinfo{author}{\bibfnamefont{V.~M.} \bibnamefont{Abazov}}
  \bibnamefont{et~al.} (\bibinfo{collaboration}{D0}),
  \bibinfo{journal}{Phys.Rev.} \textbf{\bibinfo{volume}{D82}},
  \bibinfo{pages}{032001} (\bibinfo{year}{2010}), \eprint{1005.2757}.

\bibitem[{\citenamefont{Mulders}(2011)}]{ttspec}
\bibinfo{author}{\bibfnamefont{M.}~\bibnamefont{Mulders}},
  \bibinfo{journal}{EPS, Grenoble, France, July 27}  (\bibinfo{year}{2011}).

\bibitem[{\citenamefont{Peskin and Takeuchi}(1992)}]{Pesk:91}
\bibinfo{author}{\bibfnamefont{M.~E.} \bibnamefont{Peskin}} \bibnamefont{and}
  \bibinfo{author}{\bibfnamefont{T.}~\bibnamefont{Takeuchi}},
  \bibinfo{journal}{Phys.Rev.} \textbf{\bibinfo{volume}{D46}},
  \bibinfo{pages}{381} (\bibinfo{year}{1992}).

\bibitem[{\citenamefont{Golden and Randall}(1991)}]{Golden:1990ig}
\bibinfo{author}{\bibfnamefont{M.}~\bibnamefont{Golden}} \bibnamefont{and}
  \bibinfo{author}{\bibfnamefont{L.}~\bibnamefont{Randall}},
  \bibinfo{journal}{Nucl.Phys.} \textbf{\bibinfo{volume}{B361}},
  \bibinfo{pages}{3} (\bibinfo{year}{1991}).

\bibitem[{\citenamefont{Holdom and Terning}(1990)}]{Holdom:1990tc}
\bibinfo{author}{\bibfnamefont{B.}~\bibnamefont{Holdom}} \bibnamefont{and}
  \bibinfo{author}{\bibfnamefont{J.}~\bibnamefont{Terning}},
  \bibinfo{journal}{Phys.Lett.} \textbf{\bibinfo{volume}{B247}},
  \bibinfo{pages}{88} (\bibinfo{year}{1990}).

\bibitem[{\citenamefont{Maksymyk et~al.}(1994)\citenamefont{Maksymyk, Burgess,
  and London}}]{Mak:93}
\bibinfo{author}{\bibfnamefont{I.}~\bibnamefont{Maksymyk}},
  \bibinfo{author}{\bibfnamefont{C.}~\bibnamefont{Burgess}}, \bibnamefont{and}
  \bibinfo{author}{\bibfnamefont{D.}~\bibnamefont{London}},
  \bibinfo{journal}{Phys.Rev.} \textbf{\bibinfo{volume}{D50}},
  \bibinfo{pages}{529} (\bibinfo{year}{1994}), \eprint{hep-ph/9306267}.

\bibitem[{\citenamefont{Burgess et~al.}(2009)\citenamefont{Burgess, Trott, and
  Zuberi}}]{Bur:09}
\bibinfo{author}{\bibfnamefont{C.}~\bibnamefont{Burgess}},
  \bibinfo{author}{\bibfnamefont{M.}~\bibnamefont{Trott}}, \bibnamefont{and}
  \bibinfo{author}{\bibfnamefont{S.}~\bibnamefont{Zuberi}},
  \bibinfo{journal}{JHEP} \textbf{\bibinfo{volume}{0909}}, \bibinfo{pages}{082}
  (\bibinfo{year}{2009}), \eprint{0907.2696}.

\bibitem[{\citenamefont{del Aguila et~al.}(2010)\citenamefont{del Aguila,
  de~Blas, and Perez-Victoria}}]{del:10}
\bibinfo{author}{\bibfnamefont{F.}~\bibnamefont{del Aguila}},
  \bibinfo{author}{\bibfnamefont{J.}~\bibnamefont{de~Blas}}, \bibnamefont{and}
  \bibinfo{author}{\bibfnamefont{M.}~\bibnamefont{Perez-Victoria}},
  \bibinfo{journal}{JHEP} \textbf{\bibinfo{volume}{1009}}, \bibinfo{pages}{033}
  (\bibinfo{year}{2010}), \eprint{1005.3998}.

\bibitem[{\citenamefont{Grinstein and Wise}(1991)}]{Grin:91}
\bibinfo{author}{\bibfnamefont{B.}~\bibnamefont{Grinstein}} \bibnamefont{and}
  \bibinfo{author}{\bibfnamefont{M.~B.} \bibnamefont{Wise}},
  \bibinfo{journal}{Phys.Lett.} \textbf{\bibinfo{volume}{B265}},
  \bibinfo{pages}{326} (\bibinfo{year}{1991}).

\bibitem[{\citenamefont{Chen et~al.}(2009)\citenamefont{Chen, Cline, and
  Frey}}]{Chen:2009ab}
\bibinfo{author}{\bibfnamefont{F.}~\bibnamefont{Chen}},
  \bibinfo{author}{\bibfnamefont{J.~M.} \bibnamefont{Cline}}, \bibnamefont{and}
  \bibinfo{author}{\bibfnamefont{A.~R.} \bibnamefont{Frey}},
  \bibinfo{journal}{Phys.Rev.} \textbf{\bibinfo{volume}{D80}},
  \bibinfo{pages}{083516} (\bibinfo{year}{2009}), \eprint{0907.4746}.

\bibitem[{\citenamefont{Heeck and Rodejohann}(2011)}]{Heeck:2011md}
\bibinfo{author}{\bibfnamefont{J.}~\bibnamefont{Heeck}} \bibnamefont{and}
  \bibinfo{author}{\bibfnamefont{W.}~\bibnamefont{Rodejohann}}
  (\bibinfo{year}{2011}), \eprint{1109.1508}.

\bibitem[{\citenamefont{Amsler et~al.}(2008)}]{PDG:08}
\bibinfo{author}{\bibfnamefont{C.}~\bibnamefont{Amsler}} \bibnamefont{et~al.}
  (\bibinfo{collaboration}{Particle Data Group}), \bibinfo{journal}{Phys.Lett.}
  \textbf{\bibinfo{volume}{B667}}, \bibinfo{pages}{1} (\bibinfo{year}{2008}).

\end{thebibliography}

\end{document}